\documentclass[aps,pra,prabib,twocolumn,showpacs,nofootinbib]{revtex4}
\usepackage{graphicx} \usepackage{amsmath} \usepackage{amssymb}
\usepackage{amsfonts} \usepackage{bm}
\usepackage{hyperref}

\begin{document}

\newcommand{\be}{\begin{equation}} \newcommand{\ee}{\end{equation}}
\newcommand{\bea}{\begin{eqnarray}}\newcommand{\eea}{\end{eqnarray}}


\title{Localization at threshold in noncommutative space}

\author{\href{http://pulakgiri.googlepages.com/home}{Pulak Ranjan Giri}} 
\email{pulakranjan.giri@saha.ac.in}

\affiliation{Theory Division, Saha Institute of Nuclear Physics,
1/AF Bidhannagar, Calcutta 700064, India}

\begin{abstract}
The ground state energy of a scale  symmetric system usually does
not possess any lower bound, thus making the system quantum
mechanically unstable. Self-adjointness and renormalization techniques 
usually provide the system a scale and thus
making the ground state bounded from below. We on the other hand use
noncommutative quantum mechanics and exploit the noncommutative
parameter $\Theta$ as a scale for a  scale symmetric system. The
resulting Hamiltonian for the system then allows an unusual bound
state at the threshold of the energy, $E=0$. Apart
from the Hamiltonian $\widehat{H}$ we also compute the other two
generators of the $so(2,1)$ algebra, 
the dilation $\widehat{D}$ and the  conformal
generator $\widehat{K}$ in the noncommutative space. The $so(2,1)$
algebra is not closed in the noncommutative space, but the limit
$\Theta\to 0$ smoothly goes to the $so(2,1)$ algebra restoring the
conformal symmetry. We also discuss the system for large
noncommutative parameter.
\end{abstract}


\date{\today}

\maketitle
The study of physics in noncommutative spacetime
\cite{michael,calmet} or only in noncommutative space
\cite{gamboa,gamboa1,muthu} has  become an independent field of
research work for a long time. It  started with the work of Snyder
\cite{snyder,snyder1}, where Electromagnetic theory is considered in
noncommutative spacetime. It is a well known fact that the
coordinates of a plane become noncommutative when the quantum
mechanical system in a magnetic field (perpendicular to plane)
background  is  confined in lowest Landau level. However
non-commutativity was  present in theoretical
physics before the concept of noncommutativity in spacetime or in
space coordinates was introduced. For example, the canonically
conjugate operators like coordinate $x^i$ and its conjugate momenta
$p^i$ are noncommutative $\left[x^i,p^i\right]=i\hbar$, which leads
to the uncertainty principle $\Delta x^i\Delta p^i\geq \hbar/2$ in
quantum mechanics. On the other hand although the different momentum
components do commute, it is known that the components of a
generalized momenta in the background magnetic field,
$\boldsymbol{B}= (B^1, B^2, B^3)$, do not commute,
$\left[P^i,P^j\right]=i\epsilon^{ijk}B^k$.

Noncommutativity and its effect is studied in diverse fields
starting from Quantum Field Theory
\cite{madore,calmet1,rba,bal1,falomir1,lopez1,car1}, String theory
\cite{seiberg,doug,chu1,chu2} to quantum mechanics
\cite{hov1,hov2,hov3,hov4,hov5,branko,gieres,
nair,jonke,rban,rban1,rban2,gam,chai1,bellucci3,
bellucci4,barbosa,daval,kara,dadic,jella1,jella2,dayi,dayi1}. In
quantum mechanical context several models are studied in
noncommutative space. The list includes harmonic oscillator
\cite{agni}, Hydrogen atom problem \cite{bellucci2,chai}, Zeeman
effect and Stark \cite{bellucci2} effect. Even the effect of
noncommutative space is studied for a general central potential and
solutions are obtained in large noncommutative limit \cite{gamboa}. 
It is known that usually the inclusion of spacetime 
noncommutativity  destroys the uniterity of a system but that can be restored
by a different formulation of noncommutativity of  Doplicher et at 
\cite{dopli1,dopli2}

The present letter is concerned with a scale invariant system in
non-commutative space. In particular we consider a particle on a
plane ($2D$) with an inverse square potential
\cite{giri,giri1,giri2,giri3,giri4,camblong1,camblong2,bawin}. The
importance of the inverse square potential in theoretical physics
can be understood from the huge research works carried out so far,
which in some stage can be described by an inverse square potential.
Its presence is investigated in detail in molecular physics \cite{giri4}, 
atomic physics \cite{giri2,giri4},
black hole physics  and mathematical physics. 
It shows that inverse square potential
possesses bound state solution due to the scaling anomaly caused by
quantization. Usually a length scale is introduced by 
a technique called self-adjoint extensions
\cite{reed,kumar1,kumar2,kumar3,kumar4,kumar5,kumar6,kumar7,feher}
or by renormalization \cite{rajeev}.

In this article we consider that the  space is noncommutative. This allows us 
to exploit the noncommutativity as a scale and then to study the
system. The scale symmetry of the system is thus explicitly violated
by the noncommutative parameter, $\Theta$, dependent terms in the
Taylor series expansion of the Hamiltonian. Considering inverse
square potential in the non-commutative space results to a new
problem, which falls under an interesting class of potentials
$V_\mu=\mbox{g}/r^\mu$, $\mu>2$ \cite{mako}. It is known that the system
with a potential $V_\mu$ possesses bound state solutions with energy
$E=0$. According to our standard notion $E=0$ serves as a border
line between negative energy bound sates and positive energy
scattering states.  But the fact that this notion is not always true
was shown by J. Daboul and M. M. Nieto  in \cite{jamil}, where they
argued that for potentials which asymptotically goes to zero from
above $V_\mu=0$ line, may exhibit  localized states at $E=0$. 
See a comment also \cite{hojman}.

The article is organized in the following fashion: First, we
consider the well known inverse square interaction on a plane and
shown how it changes when the co-ordinates of the plane become
non-commutative. Second, we deal with localization of the system we
considered in previous discussion, at threshold of the particle
energy, $E=0$. Third, we discuss the system for large
noncommutativity, i.e., $\Theta$ large, and then we conclude.

First, we consider a quantum mechanical system on a plane with
inverse square potential $V= \alpha {\boldsymbol{r}}^{-2}$, $\alpha$
is a constant parameter. Inverse square potential is important  both
in theoretical physics and experimental physics. It may arise in
molecular physics, when the fermions are  scattered by the
vapor of polar molecules. It has application in atomic physics and
black hole physics. From theoretical point of view this potential
$V$ carries interesting properties. For example, system with
potential $V$ possesses conformal symmetry, generated by three
operators; Hamiltonian $H$, Dilatation generator $D$ and conformal
generator $K$. In  $2D$ coordinate space the representation of the
three generators are (unit used $2m=\hbar=1$)
\begin{eqnarray}
H&=&{\boldsymbol{p}}^2 +\alpha {\boldsymbol{r}}^{-2}\,,\label{pot1}\\
D&=& Ht- \left({\boldsymbol{r}}.{\boldsymbol{p}} +
{\boldsymbol{p}}.{\boldsymbol{r}}\right)/4\,,\label{pot2}\\
K&=& Ht^2- \left({\boldsymbol{r}}.{\boldsymbol{p}} +
{\boldsymbol{p}}.{\boldsymbol{r}}\right)t/2+
{\boldsymbol{r}^2}/4\,,\label{pot3}
\end{eqnarray}
which is known to form the $so(2,1)$ algebra \cite{alfaro}
\begin{eqnarray}
\left[D,H\right]= -iH,~~ \left[D,K\right]= iK,~~ \left[H,K\right]=
2iD\,.\label{algebra1}
\end{eqnarray}
It is to be noted that due to the scale symmetry of the system,
described by the Hamiltonian $H$, there is no lower bound for the
bound state of the system. This system is thus usually unstable. But
it is known that due to scaling anomaly the system may have a finite
bound state thus making the system physically meaningful from the
bound state point of view. The usual technique for providing the
system a scale are self-adjoint extension and re-normalization. We
however study the system in noncommutative space and exploit the
scale $\Theta$ involved in the noncommutative space
($\widehat{x_1},\widehat{x_2}$).
The standard commutator algebra
\begin{eqnarray}
\left[x_i,x_j\right]= 0,~\left[p_i,p_j\right]= 0,~
\left[x_i,p_j\right]=i\delta_{ij}\,,\label{algebra2}
\end{eqnarray}
defined over the phase space is modified due to the noncommutative
scale $\Theta$ as
\begin{eqnarray}
\left[\widehat{x_i},\widehat{x_j}\right]= 2i\epsilon_{ij}\Theta,~
\left[\widehat{p_i},\widehat{p_j}\right]=0~
\left[\widehat{x_i},\widehat{p_j}\right]=i\delta_{ij}\,,\label{}
\end{eqnarray}
where  $\epsilon_{12}=-\epsilon_{12}=1$,
$\epsilon_{11}=\epsilon_{22}=0$. Assuming that $\lim_{\Theta\to
0}\left[\widehat{x_i},\widehat{x_j}\right]\to \left[x_i,x_j\right]$
is defined, one can get a realization of the noncommutative
phase space coordinates
($\widehat{x_1},\widehat{x_2},\widehat{p_1},\widehat{p_2}$) in terms
of standard coordinates
\begin{eqnarray}
\nonumber \widehat{x_1}&=&x_1 -\Theta p_2,~~ \widehat{x_2}=x_2 +\Theta p_1\\
\widehat{p_1}&=&p_1,~~~~~ \widehat{p_2}=p_2 \label{}
\end{eqnarray}
The $so(2,1)$ generators are supposed to get modified, where all
coordinates and momentum in  Eqs. (\ref{pot1})- (\ref{pot3}) will
be replaced by corresponding noncommutative counterpart
($\widehat{x_1}, \widehat{x_2}, \widehat{p_1}, \widehat{p_2}$). Note
that the introduction of noncommutative coordinates remove the
singularity of the observables. Keeping up to first order term in
$\Theta$ in the Taylor series expansion the generators become
\cite{bellucci1}
\begin{eqnarray}
\widehat{H}&=& H + \Theta 2\alpha r^{-4}L_z\,
\label{HDK1} \\
\widehat{D}&=& D + \Theta 2\alpha r^{-4}L_zt\,\label{HDK2}\\
\widehat{K}&=& K + \Theta (2\alpha t^2r^{-4}-1/2)L_z\,\label{HDK3}
\end{eqnarray}
Now the new commutators, formed by the the generators by  keeping
only first order in $\Theta$ terms,
\begin{eqnarray}
\left[\widehat{D},\widehat{H}\right]&=&-i\widehat{H}- \Theta 2i\alpha r^{-4}L_z\,,\label{cm1}\\
\left[\widehat{D},\widehat{K}\right]&=&i\widehat{K}
+\Theta\left(2i\alpha t^2r^{-4}
+i/2\right)L_z\,,\label{cm2} \\
\left[\widehat{K},\widehat{H}\right]&=&-2i\widehat{D}- \Theta
4i\alpha tr^{-4}L_z\,,\label{cm3}
\end{eqnarray}
are not closed. Note that the scale symmetry is explicitly broken by
the term $\Theta 2\alpha r^{-4}L_z$ in Eq. (\ref{HDK1}). Note that
the commutative limit $\Theta\to 0$ restores the $so(2,1)$ algebra,
namely Eqs. (\ref{cm1}) - (\ref{cm3}) reduce to Eq.
(\ref{algebra1}). In the next section we discuss the bound state
property of the system at threshold.

Second, we now discuss the system described by the Hamiltonian
$\widehat{H}$. Since we are interested in the localization
properties at the threshold $E=0$, the Schr\"{o}dinger eigenvalue
equation becomes
\begin{eqnarray}
\left({\boldsymbol{p}}^2 +\alpha {\boldsymbol{r}}^{-2}+ \Theta
2\alpha r^{-4}L_z\right)\psi=0 \label{newham}
\end{eqnarray}
where $\psi$  is a simultaneous eigenstate of $\widehat{H}$ and
$L_z$, with eigenvalues  $0$ and $m$ respectively.  The $so(2)$
symmetry (rotation about $z$ axis) is intact even in noncommutative space. 
Eq. (\ref{newham})
is separable in polar coordinates ($r, \phi$). The angular
eigenfunction is $\Phi= \exp(im\phi)$, with eigenvalue equation
$L_z\Phi=m\Phi$. The ansatz $\psi= R(r)\Phi$ with a further
similarity transformation $R(r)=\chi(r)/\sqrt{r}$ reduces  Eq.
(\ref{newham}) to a $1D$ equation
\begin{eqnarray}
\left(-\frac{d^2}{dr^2}+\frac{4m^2+4\alpha-1}{4r^2}+\frac{2\Theta\alpha
m}{r^4}\right)\chi(r)=0 \label{newham1}
\end{eqnarray}
The localized solution of (\ref{newham1}) apart from normalization
constant is
\begin{eqnarray}
R(r)= J_{\sqrt{m^2+\alpha}}\left(\frac{\sqrt{-2\Theta\alpha
m}}{r}\right)\,, \label{solution}
\end{eqnarray}
where the constraint $\sqrt{m^2+\alpha} >1$ \cite{mako} needs to be
satisfied in order the solution  to be normalizable,  which is found
from the normalization constant.

Third, the solution for  generic central potential $V(r)$ for large
noncommutativity, i.e.,  $\Theta$ large, is solved in
\cite{gamboa}. Our inverse square potential is a special  case of
\cite{gamboa} for $\Theta$ large, where now $V(r)= \alpha r^{-2}$.
Now the Hamiltonian $H$ in noncommutative space becomes
\begin{eqnarray}
\widehat H= \boldsymbol{p}^2
+\alpha\left(\Theta^2{\boldsymbol{p}^2}+r^2-2\Theta L_z\right)^{-1}
\label{1largen}
\end{eqnarray}
In order to solve the system with Hamiltonian (\ref{1largen}), it is
useful to first solve the system with Hamiltonian
$H_\Theta= \Theta^2{\boldsymbol{p}^2}+r^2-2\Theta L_z$
in Schwinger representation. The annihilation operators \cite{agni}
\begin{eqnarray}
\nonumber \widehat{a_+}=(x_1-ix_2)+\Theta(ip_1+p_2)\,,\\
\widehat{a_-}=(ix_1-x_2)-\Theta(p_1+ip_2)\,,
\label{representation1}
\end{eqnarray}
satisfy the  commutation relation
\begin{eqnarray}
\left[\widehat{a_+},\widehat{a_+}^\dagger\right]=
\left[\widehat{a_-},\widehat{a_-}^\dagger\right]=4\Theta\,,
\label{representation2}
\end{eqnarray}
with all the other commutations among the creation and annihilation
operators  being zero. In this representation the number operators
\begin{eqnarray}
\widehat{n_+}= \widehat{a_+}^\dagger
\widehat{a_+}\,,~~~\widehat{n_-}= \widehat{a_-}^\dagger
\widehat{a_-}\,,\label{number}
\end{eqnarray}
satisfy the eigenvalue equation
\begin{eqnarray}
\nonumber \widehat{n_+}|n_+,n_-\rangle= n_+|n_+,n_-\rangle\,,
n_+=0,4\Theta,8\Theta,12\Theta,...\\\widehat{n_-}|n_+,n_-\rangle=
n_-|n_+,n_-\rangle\,,
n_-=0,4\Theta,8\Theta,12\Theta,...\label{number1}
\end{eqnarray}
The Hamiltonian  $H_\Theta$ in the representation $|n_+,n_-\rangle$,
\begin{eqnarray}
H_\Theta= \widehat{n_-} + 2\Theta \label{nham2}
\end{eqnarray}
satisfy the eigenvalue equation
\begin{eqnarray}
H_\Theta|n_+,n_-\rangle = E_\Theta|n_+,n_-\rangle\,,\label{nham3}
\end{eqnarray}
with $E_\Theta= n_-+2\Theta$.  The eigenvalue of the Hamiltonian
$\widehat H$ in $|n_+,n_-\rangle$ basis is
\begin{eqnarray}
\nonumber E&=& \langle n_+,n_-|\boldsymbol{p}^2|n_+,n_-\rangle
+\frac{\alpha}{n_-+2\Theta}\\&=& \frac{n_++n_-+4\Theta}{4\Theta^2}+
\frac{\alpha}{n_-+2\Theta}\label{largen}
\end{eqnarray}
Note that the limit $\Theta\to 0$ can not be taken in
(\ref{largen}). It should be noted that scale symmetry is broken for large
non-commutativity also.

Finally, we considered particle interacting with inverse square
potential in non-commutative space. The non-commutative correction
($\Theta$ small) to the Hamiltonian leads to a potential
$2\Theta\alpha r^{-4}L_z$, which together with the inverse square
potential $\alpha r^{-2}$ and the potential $-(1/4)r^{-2}$ coming
form the kinetic term in $2D$ are capable of binding particle with
energy $E=0$. We show that the $so(2,1)$ algebra in commutative
space, which is responsible for conformal symmetry for such system
is no longer closed in non-commutative space. But the limit
$\Theta\to 0$  smoothly goes to the $so(2,1)$ algebra. For large
$\Theta$ the system is solved but this time the limit $\Theta\to 0$
can not be taken.


\end{document}